\documentclass[aps,prb,twocolumn]{revtex4}
\usepackage{bm,color,amsmath,amssymb,mathrsfs,latexsym,graphicx,psfrag}
\usepackage{epsfig}

\newcommand{\be}{\begin{eqnarray}}
\newcommand{\ee}{\end{eqnarray}}

\newcommand{\sign}{\mathrm{sign}}

\newcommand{\bp}{\mathbf{p}}

\renewcommand{\bar}{\overline}
\renewcommand{\hat}{\widehat}

\newcommand{\p}{\partial}
\newcommand{\half}{\frac{1}{2}}

\renewcommand{\(}{\left(}
\renewcommand{\)}{\right)}

\renewcommand{\l}{\langle}
\renewcommand{\r}{\rangle}

\begin{document}
\title{Quantum Dualities and Quantum Anomalous Hall Phases \\with Arbitrary Large Chern Numbers}
\author{Tong Chern$^{1}$}
\date{\today}
\affiliation{$^{1}$ School of Science, East China University of Technology, Nanchang 330013, China}

\begin{abstract}
{Quantum duality is a far reaching concept in contemporary theoretical physics.
In the present paper, we reveal the quantum dualities in quantum anomalous Hall (QAH) phases
through concrete two bands Hamiltonian models.
Our models can realize QAH phases
with arbitrary large Chern numbers.
In real materials these models may be realized by stacked $n$ layer systems of $c_1=1$ QAH insulators.
The topological phase transitions that can change the Chern numbers are studied.
And we investigate the gapless edge modes of our models in details,
and find a new mechanism for the bulk boundary correspondence.}
\end{abstract}
\maketitle
\bigskip

\section{Introduction}

Quantum duality is a profound and far reaching concept in contemporary theoretical physics,
which origins in statistical mechanics and electromagnetism\cite{duality1}.
The emergence of quantum duality in several areas of modern physics,
ranging from supersymmetrical gauge theories\cite{duality2}
to superstring theory \cite{duality3},
represents a major development in our understanding to quantum field theory and to superstring theory.
Searching for similar quantum dualities in condensed matter systems with topological phases\cite{TI1,TI2,TI3},
especially in the systems with quantum anomalous Hall phases\cite{QAr,qare,qa3d},
will be an interesting challenge.

A quantum anomalous Hall (QAH) phase
is the quantum Hall phase of some ferromagnetic insulators,
which can have quantized Hall conductance (characterized
by the first Chern number $c_1$), contributed
by dissipationless chiral states at sample edges, even without external magnetic field\cite{haldane}.
Recently, the QAH effect has been experimentally discovered in magnetic topological insulator of $\mathrm{Cr}$-doped
$\mathrm{(Bi,Sb)_2Te_3}$, where the $c_1=1$ phase has been reached\cite{qa}.
Search for QAH phases with higher Chern numbers could be important both for fundamental and practical interests, since the QAH effect with higher
Chern numbers can lower the contact resistance, and significantly improve the performance of the interconnect devices.
In recent years, there are many works in this direction\cite{hc00,hc0,hc01,hc1,hc2,hc3,hc4}.
Despite so much progress, the QAH phases that can have arbitrary large Chern numbers are still unknown.
To construct some concrete Hamiltonian models that can realize this kind of phases may be
a critical step to experimentally realize them.

In the present paper, we will reveal the quantum dualities in QAH phases
through concrete two bands theoretical models.
In real materials these models may be realized by stacked $n$ layer systems of $c_1=1$ QAH insulators.
Our models can realize the QAH phases with arbitrary large Chern numbers,
while in the previous works in searching for large Chern number QAH phases the Chern number is ordinarily bounded above. In the $c_1=1$ case, our models
become the model introduced in \cite{qam}.
But our new models can reveal some novel phenomenons of the QAH phases that can not be seen
in the $c_1=1$ case. Firstly, our models reveal a kind of quantum duality of the QAH phases.
In this duality, the characterizing parameter $g=t/m$ of a phase is equivalent to its inverse $g^{-1}=m/t$.
More specifically, under this quantum duality a model with parameter $g=t/m$
is mapped to a dual model with $g'=1/g$, which is describing the same phase.
Secondly, our models reveal a kind of peculiar quantum self duality of the QAH phases.
In this self-duality,
a model remains in the same topological phase under the self dual transformation $w(p)\rightarrow 1/w(p)$
(where $w(p)$ is a characterizing function of the topological phase).
Moreover, our models exhibit a new mechanism for the bulk boundary correspondence, which is more subtle than the usual mechanism in the $c_1=1$ case.
Finally, in our models, the novel topological phase transitions that can change the Chern numbers can be theoretically studied in details.

The present paper is organized as follows. In section 2,
we construct a large class of models for QAH phases, in these models,
the novel quantum dualities in QAH phases can be revealed.
In this section, we also analytically derive a simple formula for the Chern numbers of our models,
by utilizing a homotopical argument.
And we will demonstrate that our models can realize the QAH phases with arbitrary large Chern numbers.
The topological phase transitions that change the Chern numbers are also studied in details.
Section 3 study the gapless edge modes for the large Chern number phases.
Where, a new mechanism for bulk boundary correspondence will be revealed.
In the last section, we discuss some possible further developments and applications of our theory.

\section{Quantum Dualities, Large Chern Numbers, and Topological Phase Transitions}

In this section we will reveal the remarkable quantum dualities in QAH phases by constructing explicit Hamiltonian models.
And we will demonstrate that our models can realize the QAH phases with arbitrary large Chern numbers.
A simple formula for the Chern numbers of our models will be analytically derived by utilizing a homotopical argument.
And the intriguing topological phase transitions between the QAH phases with different Chern numbers will also be investigated.

Let's consider a class of two bands continuous models for QAH phases,
with the Bloch Hamiltonian given by
\be H(\bp)=
\(\begin{array}{cc}
m+t|w(p)|^2 & \bar{w}(p)\\
w(p) & -(m+t|w(p)|^2)
\end{array}\),\label{md3}\ee
where $t$ and $m$ are two real parameters, $p=p_x+ip_y$ is the complex momentum.
And the characterizing function $w(p)$ is a holomorphic function of $p$.

As we will prove later, this model $H(\bp)$ (\ref{md3}) is dual to the model $H'(\bp)$ given by
\be H'(\bp)=
\(\begin{array}{cc}
t+m|w(p)|^2 & {w}(p)\\
\bar{w}(p) & -(t+m|w(p)|^2)
\end{array}\).\label{md4}\ee
In this duality, the $w(p)$ in the original model $H(\bp)$ is replaced by the $\bar{w}(p)$ in $H'(\bp)$,
and at the same time the two real parameters $m$ and $t$ are switched.
Therefore, under this duality the characterizing parameter $g=t/m$ of $H(\bp)$ is mapped to $g'=1/g=m/t$ in the dual model $H'(\bp)$.
These two models $H(\bp)$ and $H'(\bp)$ are dual to each others since they always describe the same topological phase (as we will proved).
This duality is reminiscent of the Abelian duality of two dimensional quantum field \cite{chernd}.

It is easy to find the energy bands of (\ref{md3}), with
\be E(\bp)_{\pm}=\pm\sqrt{(t|w|^2+m)^2+|w|^2}.\ee
If $m\neq 0$, the band gap will be open on the whole momentum space.
When $m\cdot t>0$, the system is topologically trivial, since in this case we can continuously deform the characterizing function $w(p)$ to zero without closing the band gap.
But when $m\cdot t<0$, the system will transit to a topologically nontrivial phase which can not be deformed
to the $m\cdot t>0$ phase without closing the bulk gap.
In fact, in this case if we take $t>0$, $H(\bp)$ (\ref{md3}) will be homotopically equivalent to
\be
\(\begin{array}{cc}
|w(p)|^2-1 & 2\bar{w}(p)\\
2w(p) & -(|w(p)|^2-1)
\end{array}\).\label{md3sol}\ee

The wave function $\Psi(p_x, p_y)$ of the occupied band of (\ref{md3sol})
can be easily carried out,
\be\Psi(p_x, p_y)=\(\begin{array}{cc}
 1\\
-w(p)
\end{array}\)/\sqrt{1+|w(p)|^2}.\ee
And one can easily calculate the Berry phase ${\mathcal{A}}=i\Psi^{\dagger}d\Psi$ and Berry curvature $\mathcal{F}=d{\mathcal{A}}$.
Consequently, one can get a simple formula for the first Chern numbers of our models (\ref{md3}),
\be c_1=\frac{\sign(t)-\sign(m)}{2}\cdot\frac{i}{2\pi}\int_{\bp}\frac{dw\wedge d\bar{w}}{(1+|w|^2)^2}.\label{e}\ee
Here we are integrating over the whole momentum space,
and we have assigned $c_1=0$ for the topologically trivial phase at $m\cdot t>0$.

By this Chern number formula (\ref{e}), we can easily see that our pair of models (\ref{md3}) and (\ref{md4})
always have the same Chern numbers (since we switch $t,m$ and $w, \bar{w}$ at the same time, with $t \leftrightarrow m$ $w \leftrightarrow\bar{w}$),
hence they are pair of dual models that are describing the same topological phase.

\begin{figure} [h]
\begin{center}
\includegraphics[width=3cm]{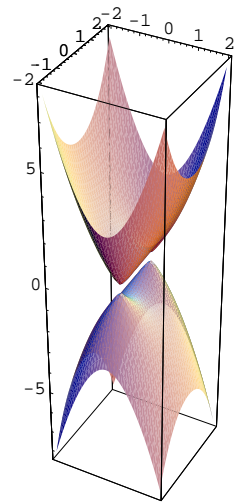}
\caption{Conduction and valence bands of our sample model with $t=0.01,m=-0.2$ are plotted as functions of momentum.} \label{two}
\end{center}
\end{figure}

To give more explicit constructions for our models (\ref{md3}), we now take the holomorphic function $w(p)$ as
\be w(p)=A\prod^n_i(p-a_i)=A\prod^n_i(p_x+ip_y-a_i),\label{w}\ee where $A$ and $a_i, i=1,...,n$ are all complex parameters.
One can use our Chern number formula (\ref{e}) to calculate the Chern numbers of this kind of models (\ref{w}).
The result is simply
\be c_1=\frac{\sign(t)-\sign(m)}{2}\cdot n.\ee
For example, one can consider a particular model
with $w(p)=p^2+1=(p+i)(p-i)$(Fig.\ref{two}), in the topologically nontrivial phase,
this model has the Chern numbe $c_1=2$.
But since in our construction of (\ref{w}) the integer $n$ is arbitrary and can be arbitrary large, hence our models with the Hamiltonian given by (\ref{md3})
and the characterizing holomorphic function $w(p)$ given by (\ref{w}) can realize QAH phases with arbitrary large Chern numbers.

Therefore, to any given large integer $n$, we have constructed a concrete QAH effect model,
with the Hall conductance $\sigma_H=-c_1e^2/h=-ne^2/h$\cite{TKNN}.
In real materials this model may be realized by stacked $n$ layer systems of $c_1=1$ QAH insulators.
And it is easily to see that if $n=1$, the holomorphic function $w(p)$ in our models
will become $w(p)=A(p_x+ip_y)$, and $|w|^2\propto \bp^2$,
hence our model (\ref{md3}) will become the model of \cite{qam}.

To study the topological phase transitions that can change the Chern numbers,
and to study the peculiar quantum self duality of the QAH phases.
We now allow the characterizing function $w(p)$ can be a meromorphic function.
And for simplicity, we further require $w(p)\rightarrow A$ when $p\rightarrow \infty$, where $A$ is a complex constant.
Hence $w(p)$ can be explicitly written as
\be w(p)=A\prod^n_i\frac{p-a_i}{p-b_i}=A\prod^n_i\frac{p_x+ip_y-a_i}{p_x+ip_y-b_i},\label{merow}\ee
where $a_i$ are zeros of $w(p)$, and $b_i$ are poles of $w(p)$,
all these zeros and poles are located at different positions of the complex $p$ plane.
Each zero $a_i$ can be paired with a pole $b_i$,
hence our model is characterized by these $n$ zero-pole pairs $\{a_i, b_i\}$.

In our present constructions (\ref{merow}), by using the Chern number formula (\ref{e}),  one can also get $c_1= n\cdot(\sign(t)-\sign(m))/2$.
This can be seen even without explicit computation since it is clear from (\ref{merow}) that the inverse function $p=p(w)$ is $n-$valued.
This implies that the $w-$Riemann sphere is covered $n$ times by the $p-$Riemann sphere.
That is to say, in the topologically nontrivial phase, each zero-pole pair will contribute a $1$ to the Chern numbers.

\begin{figure} [h]
\begin{center}
\includegraphics[width=5cm]{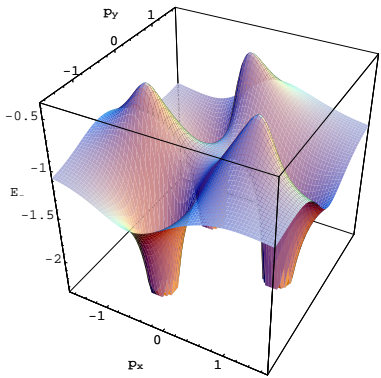}
\caption{Valence band $E(\bp)_-$ of our sample model with $t=0.01,m=-0.4$ is plotted as function of momentum.} \label{j}
\end{center}
\end{figure}

For example, one can consider a particular model with three zero-pole pairs,
with $w(p)=(p-i)(p-i\omega)(p-i\omega^2)/((p-1)(p-\omega)(p-\omega^2))$(Fig.\ref{j})(where $\omega=-\half+\frac{\sqrt{3}}{2}i$), in the topologically nontrivial phase,
this model has the Chern numbe $c_1=3$.

Now if we continuously deform the positions of zeros and poles of the constructions (\ref{merow}),
to let some zero $a_i$ coincide with some pole $b_j$, with $a_i=b_j$, then this zero will annihilate with this pole (as one can easily see from (\ref{merow})).
In the topologically nontrivial phase, this zero-pole annihilation will give rise to
a topological phase transition which will reduce the Chern numbers by $1$.

Conversely, a zero-pole pair may be created by some perturbations.
This zero-pole creation will also give rise to a topological phase transition which will increase the Chern numbers by $1$.

We now turn to the quantum self duality of the QAH phases that can be revealed by our models. We first notice that,
in the constructions of (\ref{merow}), the statuses of zeros and poles are equal.
Thus our models are describing the same topological phases under the zero-pole switching.
That implies that the same topological phase can be characterized
by zero-pole pairs $\{a_i, b_i\}$ or by the switched pairs $\{b_i, a_i\}$.
We will call this zero-pole duality of the QAH phases as the quantum self-duality.

This quantum self-duality can also be understood from the Chern number formula (\ref{e}).
This formula has the peculiar property that it is invariant under the self dual transformation
\be w\rightarrow \frac{1}{w}. \ee
This indicates that the QAH phases are self dual under $w(p)\rightarrow 1/w(p)$.
Of course, this self duality switches the zeros and poles of the characterizing meromorphic function $w(p)$.

\section{Novel Mechanism for Bulk Boundary Correspondence}

In this section, we will study the gapless edge modes of the models
(for QAH phases with arbitrary Chern numbers) that we proposed in last section.
This study will reveal a new mechanism for bulk boundary correspondence,
which can not be seen in the $c_1=1$ case.

For simplicity, we will take the model with the characterizing function $w(p)=Ap^2$ as an example.
This model is homotopically equivalent (if we take $t>0$) to the Hamiltonian $H(p_x,p_y)$ given by
\be \(\begin{array}{cc}
m+|A|^2(p^2_x+p^2_y)^2 & 2\bar{A}(p_x-ip_y)^2\\
2A(p_x+ip_y)^2 & -(m+|A|^2(p^2_x+p^2_y)^2)
\end{array}\).\label{hotm1}\ee
Hence we will focus on the model with this specific Hamiltonian (\ref{hotm1}),
but our conclusions can be applied to the whole homotopically class.

As we have discussed in last section, if $m>0$ this model (\ref{hotm1}) will be in topologically trivial phase,
while when $m<0$, it will be in a nontrivial phase with $c_1=2$. In the $m<0$ phase, the bulk low energy effective theory of (\ref{hotm1}) is a Chern-Simons theory at level $k=c_1=2$\cite{pt1}, this theory is anomaly at the spatial boundary, this anomaly should be cancelled by the chiral anomaly of some gapless chiral edge modes\cite{witten1,witten2}. This bulk boundary correspondence tells us that in the $m<0$ topologically nontrivial phase the numbers of chiral edge modes in our model (\ref{hotm1}) should be $2$, while in the $m>0$ phase, this number should be $0$. We will show that this is exactly the case, but the detail mechanism is quite different with the $c_1=1$ case\cite{QAr}.

In what follows, we will investigate the gapless edge modes and the detail mechanism for bulk boundary correspondence of our model (\ref{hotm1}),
by solving this model in half space with an open boundary condition.

Assuming that we are solving the model (\ref{hotm1}) in half space with $x\geq 0$. Hence $p_y$ remains a good quantum number,
but the $p_x$ in (\ref{hotm1}) should be replaced by the operator $\hat{p}_x=-i\p/\p x$.
Now, let's firstly consider the case of $p_y=0$. From (\ref{hotm1}) we can see that in this case
the corresponding Hamilton operator $\hat{H}_0$ should be
\be \hat{H}_0=\(\begin{array}{cc}
m+|A|^2{\p^4}/{\p x^4} & -2\bar{A}{\p^2}/{\p x^2}\\
-2A{\p^2}/{\p x^2} & -(m+|A|^2{\p^4}/{\p x^4})
\end{array}\).\ee
We will search for the normalizable zero energy solution (zero mode) $\psi(x)$ that satisfies $\hat{H}_0\psi(x)=0$.
The requirement of normalizability will impose a boundary condition, $\psi(\infty)=0$.
Further more, $\hat{p}_x$ should be a Hermitian operator, this will impose another boundary condition, $\psi(0)=0$.

Assuming that the zero mode $\psi(x)$ can be written as a linear superposition of
the functions in forms $e^{\lambda x}\phi$,
where $\phi$ is a two components constant vector.
By using the zero mode equation we can have
\be\(m+|A|^2\lambda^4-2i\lambda^2|A|
\tau_{\theta}\)\phi=0,\label{phieq}\ee
where we have rewritten the complex parameter $A$ as $A=|A|e^{i\theta}$, and $\tau_{\theta}$ is a $2\times 2$ Hermitian matrix $\tau_{\theta}=\(\begin{array}{cc}
0 & -ie^{-i\theta}\\
ie^{i\theta} & 0
\end{array}\)$.
We will denote the eigenvalues of $\tau_{\theta}$ as $s$, obviously $s=\pm 1$.

As one can easily see that, equation (\ref{phieq}) means $\phi$ must be one of the eigenstates of
$\tau_{\theta}$ with eigenvalue $s$, denoted as $\phi_{s}$, with $\tau_{\theta}\phi_{s}=s\phi_{s}$.
Consequently, equation (\ref{phieq}) becomes
\be|A|^2\lambda^4-2i|A|s\lambda^2+m=0.\label{ll}\ee
The roots of this equation are $(\lambda^2)_{\pm}=i(s\pm\sqrt{1+m})/|A|$.
It is easy (by (\ref{ll})) to see that, the roots for the case of $s=-1$ are the complex conjugations of the roots for $s=1$.
For any one of the two values of $s$, we have four roots for $\lambda$, two of these roots have positive real parts, and the other two have negative real parts. Only the roots with negative real parts can satisfy the boundary condition $\psi(\infty)=0$. We will denote these two roots as $\lambda_1, \lambda_2$. Moreover, to satisfy the boundary condition $\psi(0)=0$ at the same time, the zero mode $\psi(x)$ must take the form of $\psi_s(x)$, with
\be\psi_s(x)=(e^{\lambda_1x}-e^{\lambda_2x})\phi_{s}.\ee

Now let's consider the situation of $p_y\neq 0$. We will take $p_y$ as a small perturbation parameter.
Obviously, to satisfy the zero order (of $p_y$) equation $\hat{H}_0\psi(x)=0$, the wave functions of edge modes must be
$\psi_{p_y,s}(x,y)=\psi_s(x)e^{ip_yy}$.

Further more, from the full Hamiltonian (\ref{hotm1}),
we can read off the first order perturbation $H'$
\be H'=4\(\begin{array}{cc}
0 & -i\bar{A}\\
iA & 0
\end{array}\)p_y\hat{p_x}=4|A|\tau_{\theta}p_y\hat{p}_x.\ee
In what follows we will calculate the first order correction $\l\psi_{p_y,s}|H'|\psi_{p_y,s}\r$
to the dispersion relation of the edge modes (noticing that $\l\psi_{p_y,\pm 1}|H'|\psi_{p_y,\mp 1}\r\propto \phi^{\dagger}_{\pm 1}\tau_{\theta}\phi_{\mp 1}=0$).

In the situation of $m>0$, simple analysis tells us, whatever $s=1$ or $s=-1$,
the two relevant roots $\lambda_1, \lambda_2$ are always proportional to
$\lambda_1\varpropto -\cos(\pi/4)+i\sin(\pi/4)$, $\lambda_2\varpropto -\cos(\pi/4)-i\sin(\pi/4)$ respectively.
A few calculations show us, in this situation, $\l\psi_{s}|\hat{p}_x|\psi_{s}\r=0$.
Hence, when $m>0$, although there are gapless excitations localized at the boundary,
but in this situation the dispersion relation of these gapless edge modes must be in form of $\varepsilon(p^2_y)$
(since the first order correction is zero). Thus they can not be chiral modes, just as required by the bulk boundary correspondence.

In the situation of $m<0$, detail analysis shows us, for $s=1$, both the two relevant roots $\lambda_1, \lambda_2$ are at
the lower half of the complex plane, thus have the forms of
\be\lambda_1=-a-bi, \lambda_2=-c-di,\label{l1l2}\ee
where $a,b,c,d$ are all positive numbers. In this case, detail calculations show us $\l\psi_{s}|\hat{p}_x|\psi_{s}\r$ is a negative number,
proportional to $-(bc+ad)$. And since $\phi^{\dagger}_{s=1}\tau_{\theta}\phi_{s=1}=1$, hence $\l\psi_{p_y,s}|H'|\psi_{p_y,s}\r$ is proportional to $-p_y$.
Therefore, in the case of $s=1$, there is a left moving chiral edge mode, with the dispersion relation
\be\varepsilon=-v_Fp_y.\label{disper}\ee
 For the case of $s=-1$, detail analysis shows us, both the two relevant roots $\lambda_1, \lambda_2$ are at the upper
 half of the complex plane. These two roots are the complex conjugations of (\ref{l1l2}), thus have the forms of
$\lambda_1=-a+bi, \lambda_2=-c+di$.
Detail calculation shows us $\l\psi_{s}|\hat{p}_x|\psi_{s}\r$ is now positive,
in fact it is proportional to $(bc+ad)$. And since $\phi^{\dagger}_{s=-1}\tau_{\theta}\phi_{s=-1}=-1$,
hence we also have $\l\psi_{p_y,s}|H'|\psi_{p_y,s}\r\propto -p_y$. Consequently, we have another left moving chiral edge mode,
also with the dispersion relation (\ref{disper}).
In summary, although the detail mechanism is subtle, but in the $m<0$ topologically nontrivial phase,
we have exactly two chiral edge modes,
just as required by the bulk boundary correspondence.\\

\section{Conclusion and Discussions}

In this paper, the profound quantum dualities for QAH phases have been studied through concrete Hamiltonian models.
In one kind of these quantum dualities, a model with
characterizing parameter $g=t/m$ is dual to a model with $g'=1/g=m/t$.
In another kind of dualities, the QAH phases are self dual under $w(p)\rightarrow 1/w(p)$.
These quantum dualities are reminiscent of the quantum duality of quantum gauge theory and superstring theory,
and may shed new lights on the understanding of topological phases of quantum matters.

Moreover our models can realize the QAH phases with arbitrary large Chern numbers.
And we have studied the gapless edge states of these models, and found a new mechanism for the bulk boundary correspondence.

The studying for the chiral edge modes of quantum anomalous Hall phase with arbitrary large Chern numbers
may have potential applications in future electronic devices,
because the chiral edge modes can transport electric current without dissipations,
and the only resistances are the contact resistances,
these contact resistances can be effectively reduced by the edge modes with large Chern numbers.

We also theoretically study the topological phase transitions
between the topologically nontrivial phases with different Chern numbers.
These topological phase transitions are triggered by the annihilations or creations of the zero-pole pairs of
the characterizing meromorphic function $w(p)$.

\section{Acknowledgements}

The authors acknowledge the support of the Doctoral Startup Package Fund of East China University of Technology (No. DHBK201203).

\end{document}